\documentstyle[psbox]{WORKSHOP}
\begin{document}

% TITLE OF THE PAPER
%  If the title is too long for a single line, you can split it 
%  by putting two backslashes. 
%  You might want to put the subtitle. Then it should be inserted 
%  within {\large\sf  }.
%  e.g.:  
%     \title{ Too Long Title \\ for one line \\
%     {\large\sf Subtitle} }
\title{
SSC of MAXI experiment %\\ 
%{\large\sf  -- Brief Instructions for Users of the `WORKSHOP' Style
%File --} % SUBTITLE
}

% AUTHOR(S) 
\author{
 M. Sakano$^{1}$, H. Tomida$^1$, M. Matsuoka$^1$, S. Ueno$^1$, 
S. Komatsu$^1$, Y. Shirasaki$^1$, M. Sugizaki$^1$, \\
K. Torii$^1$, W. Yuan$^1$, E. Miyata$^{2,1}$, H. Tsunemi$^{2,1}$, T. Kamazuka$^2$, C. Natsukari$^2$,
M. Jobashi$^3$,\\
I. Tanaka$^4$,
N. Kawai$^{5,6,1}$, T. Mihara$^{6,1}$, H. Negoro$^6$ and A. Yoshida$^{7,6,1}$
\\[12pt]  % TO BE SPACED WITH ONE LINE
%
% INSTITUTES OF AUTHORS
$^1$ SURP/NASDA, 2-1-1 Sengen, Tsukuba, Ibaraki 305-8505, Japan\newline\\
$^2$ Dept. of Earth and Space Science, Osaka University, 1-1 Machikaneyama, Toyonaka, Osaka 560-0043, Japan\\
$^3$ ICRR, University of Tokyo, 5-1-5 Kashiwa-no-Ha, Kashiwa City, Chiba 277-8582, Japan\\
$^4$ The Graduate University for Advanced Studies, Hayama-chou, Miura, Kanagawa 240-0193, Japan\\
$^5$ Dept. of Physics, Tokyo Institute of Technology, 2-12-1 Ookayama, Meguro-ku, Tokyo 152-8551, Japan\\
$^6$ Cosmic Radiation Laboratory, RIKEN, 2-1 Hirosawa, Wako, Saitama 351-0198, Japan\\
$^7$ Dept. of Physics, College of Science \& Engineering, \\ Aoyama Gakuin University, 6-16-1 Chitosedai, Setagaya, Tokyo 157-8572, Japan\\
%
% please put the first author's initial and e-mail address below
{\it E-mail(MS): sakano@oasis.tksc.nasda.go.jp} 
%            \_ Initial      \
%                             \_ E-mail address
}

\abst{
  Monitor of All-sky X-ray Image (MAXI) on the International Space Station (ISS)
 has two kinds of X-ray detectors: the Gas Slit Camera (GSC) and
 the Solid-state Slit Camera (SSC).
  SSC is an X-ray CCD array, consisting of 16 chips, which
 has the best energy resolution as an X-ray all-sky monitor
 in the energy band of 0.5 to 10 keV.
  Each chip consists of 1024$\times$1024 pixels with a pixel size of 24$\mu$m,
 thus the total area is $\sim$ 200 cm$^2$.
  We have developed an engineering model of SSC, i.e., CCD chips, electronics,
 the software and so on, and have constructed the calibration system.
  We here report the current status
 of the development and the calibration of SSC.
}

\kword{X-ray: detectors --- detectors: CCD --- SSC --- International Space Station (ISS)}

\maketitle
\thispagestyle{empty}

\section{MAXI}

 Monitor of All-sky X-ray Image (MAXI: Matsuoka et al. 1997a, 1997b, 1999;
 Kawai et al. 1999; Mihara et al. 1999, 2000;
 Torii et al. 1999; Tomida et al. 2000a; 
 Mihara et al. 2001b in this proceeding)
 is one of the first payloads of  
 Japanese Experiment Module (JEM or KIBO)/Exposed Facility 
 on the International Space Station (ISS).
 In one orbit of the ISS with 90 minutes, MAXI covers the whole sky.
 It is finally expected to monitor activities of about 2000--3000 X-ray sources
 with the limiting sensitivity of up to 1~mCrab (for detail, see  
 Yuan et al. 2001 in this proceeding).
 The observation is scheduled to start in the beginning of 2005
 and continue for two years.

 MAXI consists of two kinds of X-ray detectors: the Gas Slit Camera (GSC:
 Mihara et al. 2001a, 2001b) and
 the Solid-state Slit Camera (SSC: Torii et al. 1999; Miyata et al. 1999, 2000;
 Tomida et al. 2000b; Kamazuka et al. 2001).  The X-ray detector of the SSC
 is a CCD array,
 whereas that of the GSC is gas proportional counter.  Hence, SSC has
 better energy resolution, and better detection efficiency in the
 soft energy band.   We here report on the current status of the
 development of SSC.

\section{SSC}

\subsection{Overview of SSC camera}

    MAXI has two SSC cameras
 (for the arrangement, see Mihara et al. 2001b in this proceeding),
 each of which is identical.
 Each SSC consists of an array of 16 CCD chips.
% (Fig.~ref{fig:array}).
 The whole SSC is developed by NASDA and Osaka University.
  The principle of determining positions with a slit camera is found
 in Matsuoka et al. (1997a), Tomida et al. (2000a), and Mihara et al. (2001b)
 in this proceeding.

\begin{figure*}[htbp]
\centering
\psbox[xsize=15cm,rotate=r]
{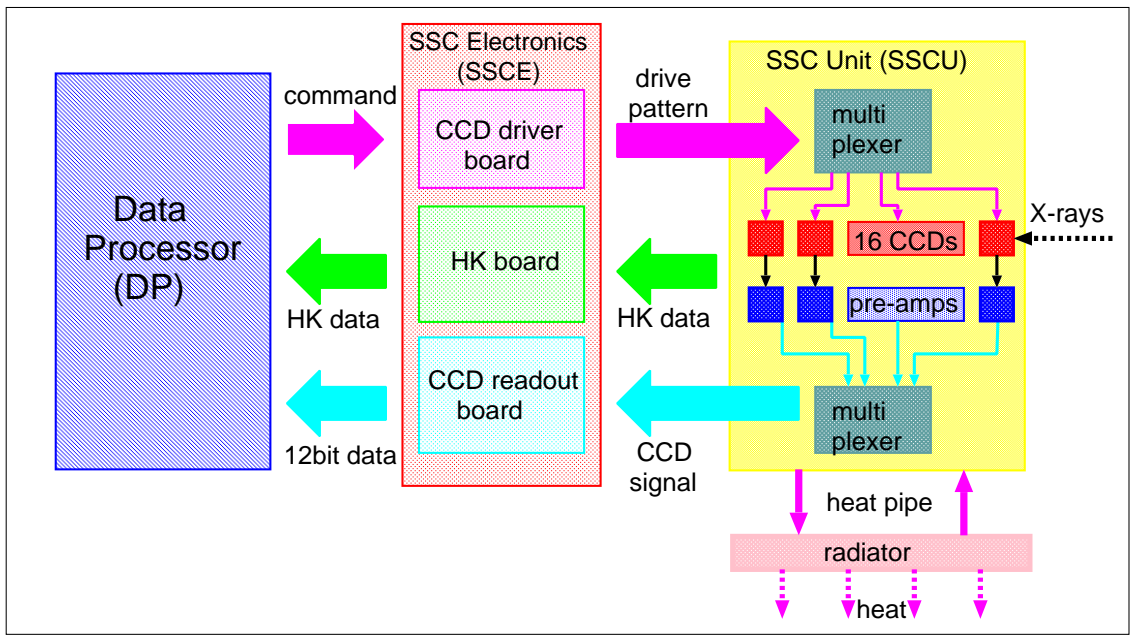}
%{electronix.eps}
\caption{Schematic view of SSC.}
\label{fig:schematic}
\end{figure*}

   The schematic view of SSC is given in Fig.~\ref{fig:schematic}.
 SSC consists of three main parts:
 SSC Unit (SSCU),
 SSC Electronics (SSCE), and Data Processor (DP).
 The CCD chips and pre-amplifiers are contained in SSCU.
 The analogue signal generated in SSCU is digitized in SSCE,
 then, handled in DP, and sent to the telemetry as data packets
 (see section~\ref{sec:data-handling}).

\begin{figure*}[htbp]
\centering
%\psbox[xsize=0.4#1,ysize=0.2#1,rotate=r]
%\psbox
\psbox[xsize=13cm]
{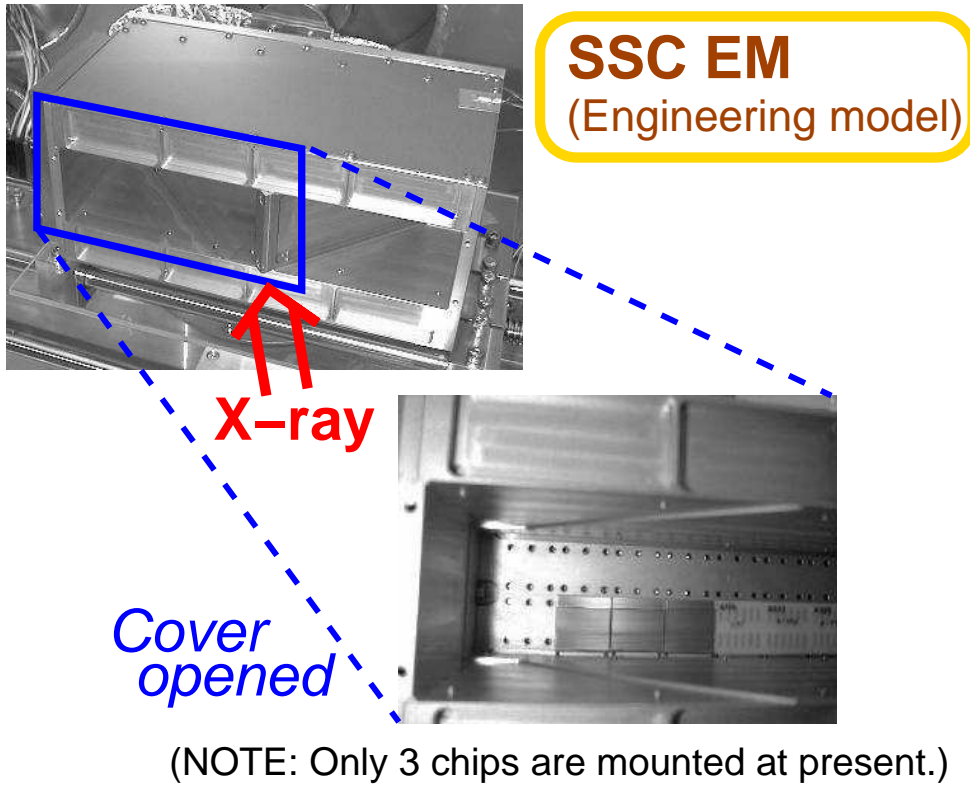}
\caption{The SSC EM where slat collimators are removed.
 Only 3 chips in 16 are currently mounted.}
\label{fig:ssc-em-photo}
\end{figure*}

\subsection{CCD}

 The chips for SSC are developed by Hamamatsu Photonics K.K. and Osaka University.
%  Fig.~\ref{fig:CCDchip} shows a picture of the MAXI-CCD,
% whereas
 Table~\ref{tbl:chip} lists the details of the CCD chip.
% Each CCD chip has three side buttable structure.
  The X-ray detection area is 25$\times$25mm$^2$ for 1-chip,
 accordingly 200~cm$^2$ in total (for 2 SSCs).
  Since SSC is a slit camera, the CCD needs only one-dimensional image;
 therefore, we adopt full-frame transfer type for the CCD chips of SSC.
  To block the optical light, aluminium (Al) is evaporated on the surface 
 of the CCD chips.  Whereas a large shock is generally expected
 in launch operation,
 this Al-coating method is tough for the shock,
 and accordingly makes it possible to design a simple
 structure for the whole camera.
  Note that this Al-coating also can minimize the radiative heat-input
 to the CCD chip.

\begin{table}
\begin{center}
\caption{details of MAXI-CCD chip \label{tbl:chip}}
\begin{tabular}{cccc}
\hline\hline\\[-6pt]
Parameters      & Present & Goal & Unit\\
\hline
pixel size      & 24 $\times$ 24 & & $\mu$m \\
pixel number    & 1024$\times$1024  & &  \\
CTE             & $>$99.999  & $>$99.999 & \%   \\
readout noise   & 6  &  $<4$ & e$^-$ (RMS) \\
depletion layer & ($^{\dag})$  & $>$40 & $\mu$m    \\
%dark current    & ??????$^{\ddag}$ & 4 & e$^-$/s/pixel \\
energy resolution & 150 & 130 & eV$^\P$ \\
\hline
clocking pattern  & \multicolumn{3}{l}{2 phase, full frame transfer} \\
front/back-side & \multicolumn{3}{l}{front-side illumination}\\
optical filter & \multicolumn{3}{l}{by Al coating} \\
\hline
temperature & \multicolumn{3}{l}{$-$60$^{\circ}$C, using buried Peltier cooler}\\
readout method  & \multicolumn{3}{l}{Integration method} \\
\hline
\end{tabular}
\label{tbl:CCD}
\end{center}
\footnotesize
\noindent
$^{\dag}$: The thickness of depletion layer of the chips mounted on EM is
 less than 20$\mu$m.  However, we have already developed the chips
 with the depletion-layer thickness of 40$\mu$m.\\
%%$^{\ddag}$: when operated at ???????$^{\circ}$C.\\
$^\P$: FWHM at 5.9~keV.
\end{table}

   X-ray CCDs must be cooled.  To do this, we combine passive radiators,
 loop heat pipes (LHP) with liquid propylene fabricated by Swales K.K.,
 and Peltier cooler buried in each CCD chip.
   The power consumed by Peltier cooler is $\sim$1 W/CCD,
 which can make the temperature difference on hot and cold sides in the
 Peltier cooler to be $\Delta T = 40^{\circ}$C.
   We expect that the hot side of the Peltier cooler can be
 cooled down to $-$20$^\circ$C by LHP and the radiator, hence
 the CCD temperature of $-$60$^\circ$C will be achieved.
   However, there may still remain some problems, which is being studied;
 e.g., the thermal environment around MAXI in the ISS is not
 completely fixed, or the condition in which the LHP works
 has not been well understood, yet.
 The detailed discussion is found in Tomida et al. (2000b).

%\subsection{Radiation Damage}

   The radiation damage is one of the most critical problems of the X-ray CCD
 observation in orbit, as seen in the case of {\it ASCA}/SIS
 (Yamashita et al. 1997) and {\it Chandra}/ACIS.
 Although the MAXI mission time of 2~years is not long, the accumulated
 radiation damage may cause significant degradation.
 We are looking for the most effective and practical method against this
 problem, such as the charge injection (CI) method (Tomida et al. 1997).
  Detailed discussion is found in Tomida et al. (2001b).

\subsection{Camera unit (SSCU), Electronics (SSCE), and Data Processor (DP) \label{sec:data-handling}}

 SSCU and SSCE are fabricated by Meisei Electric K.K.,
 and DP is, by NEC K.K.
   SSCE drives the SSCU and reads analogue signals from SSCU;
 since each SSCU has 16 CCDs, SSCE reads out the signals from those CCDs
 one after another by switching multiplexers in each SSCU. 
% (Fig.~\ref{fig:CCDread}).
 Then, SSCE digitizes the signal data into 12-bit data, and transfers them to DP.

    We have tried three readout methods (for each pixel) in SSCE:
 integration, delay, and correlated double sampling (CDS).
 The respective methods are used in {\it Chandra}/ACIS
 and {\it ASTRO-E}/XIS (Hayashida et al. 1999),
 in {\it ASCA}/SIS (Burke et al. 1994), and in commercially available (optical) CCDs.
 We recorded the best performance with the integration method,
 hence adopted it.

   The signal charges along the same columns in each chip are summed up 
 like the fast mode in {\it ASCA}/SIS (Burke et al. 1994)
 or the P-sum mode of {\it ASTRO-E}/XIS (Hayashida et al. 1999)
 because only one-dimension positional information is required.
  In default, charges of 16 pixels are summed up,
 where we can indicate the number of bins by the command.
  The readout speed is 125 kHz; accordingly the readout time for all the
 16 chips is $\sim$9 sec in the default binning.

   The algorithm of the data reduction in DP is also almost the same as that
 in the timing mode of {\it ASTRO-E}/XIS.  The DP detects events, and sends
 the position, grade, and the summed pixel-level of each event
 to the telemetry.  We have the frame- and dark-frame-modes
 for diagnosis, which are also the same as that of {\it ASTRO-E}/XIS.

\subsection{Slit and Collimators}

\begin{figure}[htbp]
\centering
%\psbox[xsize=0.4#1,ysize=0.2#1,rotate=r]
%\psbox
\psbox[xsize=8cm]
{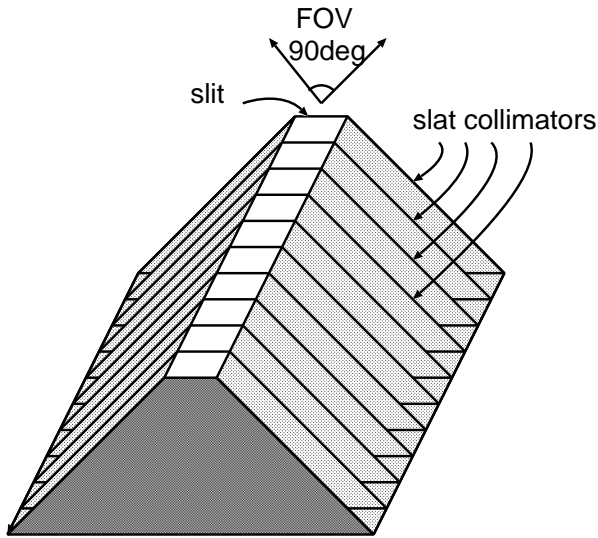}
\caption{Schematic view of the slit and the slat collimators of SSC.}
%\caption{the slit and the slat collimators}
\label{fig:collimator}
\end{figure}

  Fig.~\ref{fig:collimator} shows the schematic view of the slit and the
 slat collimator system on SSCU to determine the photon arrival direction.
  The accuracy of estimated position mainly
 depends on the size of the slit; if the slit is narrower, the accuracy is
 better, but the effective area is smaller, and vice versa.
  The field of view of each SSC is 1$\sim$2$^{\circ}$ $\times$ 90$^{\circ}$.
  The thickness of slat collimators will be $\sim$ 100$\mu$m,
 which are aligned by $\sim$3mm pitch, although these values may be changed.
  Another point for the positional accuracy is flatness and non-reflectivity
 of the collimator.  We are now searching for the best method to hold the
 flatness and to reduce surface reflectivity of the collimators.
  Mihara et al. (2001b) in this proceeding gives further information
 on this problem.
  We finally expect to determine the photon arrival direction with the accuracy
 of $\sim$1$^\circ$ or better.

\section{Calibration}

\begin{figure}[htbp]
\centering
%\psbox[xsize=0.4#1,ysize=0.2#1,rotate=r]
%\psbox
\psbox[xsize=8cm]
{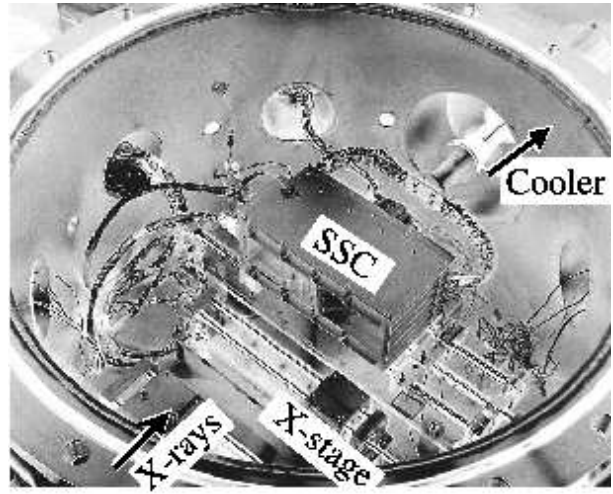}
\caption{SSC-EM is put in the center of the vacuum chamber. 
 The back-side (upper-left) of SSC is a cold plate, which is cooled
 by mechanical cooler out of vacuum chamber.  X-rays are illuminated
 from lower-left side, which is connected with another vacuum chamber where
 an X-ray generator is placed.  In addition, a radio isotope $^{55}$Fe
 can be set on the X-stage in front of SSC.}
\label{fig:SSC-vacuum}
\end{figure}

  The calibration of CCD chips and the SSC-EM (Engineering model) is
 on-going at NASDA and Osaka University.
 For the flexible calibration system at Osaka University, which aims
 in particular at the calibration in the lower energy band below 2~keV,
 see Miyata et al. (2000) and Kamazuka et al. (2001).
  For the higher energy band above 1~keV, we use fluorescent
 X-rays from various metal elements (Torii et al. 1999; Tomida et al. 2000b),
 which is similar to the calibration system for the {\it ASTRO-E}/XIS
 at Kyoto University (Hamaguchi et al. 2000).  Combining collimators and
 filters, illuminating fluorescent X-rays from each atom are purified
 according to the method in Hamaguchi et al. (2000).
 Fig.~\ref{fig:SSC-vacuum} shows the SSC-EM set in the vacuum chamber.

\begin{figure*}[htbp]
\centering
%\psbox[xsize=0.4#1,ysize=0.2#1,rotate=r]
%\psbox
\psbox[xsize=8cm]
{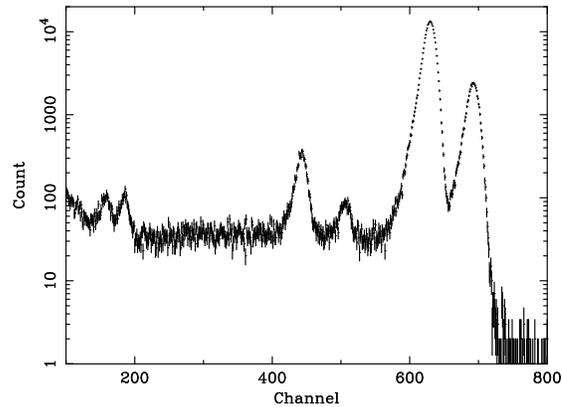}
\caption{SSC-EM spectrum of Mn K-lines ($^{55}$Fe).
  The FWHM of the main peak at 5.9~keV is 150~eV,
 where the statistical uncertainty is quite small, $\sim$0.1~eV.
}
\label{fig:fe55}
\end{figure*}

   Fig.~\ref{fig:fe55} shows the spectrum of $^{55}$Fe K-lines with
 SSC-EM in the operation at $-60 ^{\circ}$C, whereas Table~\ref{tbl:chip}
 summarizes some important parameters obtained in the current calibration.
  We are steadily recording a FWHM of 150~eV or better
 for the energy resolution at 5.9~keV where
 single pixel events are accumulated, 6~electron (RMS) for
 the read-out noise, and more than 99.999\% efficiency for the CTE
 (Charge Transfer Efficiency).  Note that these values were obtained when we
 use preliminary chips and EM electronics; the improvement is on-going.
  We have already developed
 new chips with much thicker depletion layer of more than 40$\mu$m,
 but have not combined it with the SSCE, yet.
   We have also made the preliminary version of the DP software;
 now it is being tested.

   For the calibration of the collimator, which is important to determine
 the arrival direction of each X-ray photon,  we will calibrate them
 using 19~m X-ray beam line at NASDA (see Torii et al. 1999).
  The absolute detection efficiency will be measured with well-calibrated
 reference detectors: X-ray proportional counters.
  We are now further improving all the SSCU including chips, SSCE, and DP,
 and will accomplish the SSC flight model (FM) until 2004.

\section{Summary}

   We have been developing the CCD camera, SSC, which will be onboard
 the all-sky X-ray monitor MAXI on ISS/JEM.  MAXI has two SSCs, and
 each SSC has 16 CCD chips fabricated by Hamamatsu Photonics K.K.
 At present, calibration and development of the engineering model of SSC 
 are on-going.  We achieved the energy resolution of 150~eV for the FWHM
 at 5.9~keV.  Another chips with thicker depletion layer
 of more than 40$\mu$m have been developed.  We are continuing the
 development, and plan to accomplish the flight model of SSC until 2004.
  The start of the MAXI observation is scheduled in the beginning of 2005.

\vspace{1pc}
\noindent 
The authors express their thanks to Dr. M.~J.~L. Turner
 for reviewing the manuscript.

\section*{References}

\re
Burke B. E. et al. 1994 IEEE Trans. Nuc. Sci., 41, 375
% \bibitem{Burke1994}
% B.~E. {Burke}, R.~W. {Mountain}, P.~J. {Daniels}, and V.~S. {Dolat}.
% \newblock {\em IEEE Trans. Nuc. Sci.}, 41:375, 1994.

\re
Hamaguchi K. et al. 2000 Nucl. Inst. Meth. A, 450/2-3, 360
% K. Hamaguchi, Y. Maeda, H. Matsumoto, M. Nishiuchi, H. Tomida, K. Koyama, H. Awaki, T. Tsuru
% pp.360--364

\re
Hayashida K. et al. 1998 proc of SPIE., 3445, 278
% ASTRO-E/XIS

\re
Kamazuka T. et al. 2001 proc of ``New Century of X-ray Astronomy'', submitted
%development of engineering model of the X-ray CCD camera of the MAXI experiment on the ISS/JEM
%     T.Kamazuka, C.Natsukari, E.Miyata, H.Tsunemi, H.Tomida, M.Matsuoka, S.Ueno, M.Sakano, M.Jobashi,
%     I.Tanaka 
%     New Century of X-ray Astronomy, 2001 Mar, Yokohama Symposia

%\re
%Kawai N. et al. 1996 proc of SPIE., 2808, 555
%X-ray transient monitor for JEM on the International Space Station 
%     N.Kawai, M.Matsuoka, A.Yoshida, H.Tsunemi, 
%     SPIE, 1996, 2808, 555

\re
Kawai N. et al. 1999 Astron. Nachr., 320, 372
% Monitor of All-sky X-ray Image (MAXI) for JEM on the International Space Station 
%     N.Kawai, M.Matsuoka, A.Yoshida, T.Mihara, T.Kotani, H.Negoro, Y.Shirasaki, I.Sakurai, M.Namiki, K.Torii,
%     H.Tsunemi, E.Miyata, K.Hayashida, S.Kitamoto, K.Yoshita, Y.Hashimoto, M.Yamauchi 
%     Astronomische Nachrichten, 1999, vol-320,no-4, p372

\re
Matsuoka M. et al. 1997a Proc of All-sky X-ray Observations in the Next Decade, eds. M. Matsuoka, N. Kawai (Published by RIKEN), 249

\re
Matsuoka M. et al. 1997b proc of SPIE., 3114, 414
% MAXI (monitor of all-sky x-ray image) for JEM on the Space Station 
%      M.Matsuoka, N.Kawai, T.Mihara, A.Yoshida, H.Kubo, T.Kotani, H.Negoro, B.C.Rubin, S.Shimizu, H.Tsunemi,
%      K.Hayashida, S.Kitamoto, E.Miyata, M.Yamauchi 
%      SPIE, 1997, 3114, 414

%\re
%Matsuoka M. et al. 1999 Proc of Space Technology and Applications International Forum, ed M. S. El-Genk, 458, 163
%% MAXI Mission for the Space Station 
%%     M.Matsuoka, N.Kawai, T.Mihara, A.Yoshida, H.Tsunemi, E.Miyata, K.Torii, T.Kotani, H.Kubo,
%%     H.Matsumoto, H.Negoro, B.C.Rubin, I.Sakurai, Y.Shirasaki, K.Hayashida, S.Kitamoto, M.Yamauchi, 
%%     Space Technology and Applications, International Forum, 1999, 458, p163

\re
Matsuoka M. et al. 1999 Astron. Nachr., 320, 275
%MAXI mission for space station and future ASM 
%     M.Matsuoka, M.Sugizaki, H.Tomida, K.Torii, S.Ueno, N.Kawai, T.Mihara, A.Yoshida, H.Negoro,
%     Y.Shirasaki, I.Sakurai, H.Tsunemi, E.Miyata, M.Yamauchi 
%     Astronomische Nachrochten, 1999, vol-320,no-4, p275

%\re
%Mihara T. et al. 2000 STAIF 2000,  
%Monitor of All-Sky X-Ray Image (MAXI) 
%     T.Mihara. M.Matsuoka, N.Kawai, A.Yoshida, H.Tsunemi, E.Miyata, H.Negoro, K.Torii, S.Ueno, M.Sugizaki,
%     H.Tomida, I.Sakurai, Y. Shirasaki,M. Yamauchi 
%     Space Technology and Applications, International Forum, 2000

\re
Mihara T. et al. 2000 Advances in Space Research, 25(3-4), 897
% MAXI (Monitor of All-sky X-ray Image) for JEM on the International Space Station 
%      T. Mihara, M. Matsuoka, N. Kawai, A. Yoshida, T. Kotani, H. Negoro, H. Kubo, Y. Shirasaki, B.C.Rubin,
%      K.Torii, H.Tsunemi,S.Kitamoto, K.Hayashida,E. Miyata, K.Yoshita,M.Yamauchi 
%      Advances in Space Research, 2000, vol25, issue 3-4, p897-900

\re
Mihara T. et al. 2001a proc of ``New Century of X-ray Astronomy'', submitted

\re
Mihara T. et al. 2001b in this proceeding

\re
Miyata E. et al. 1999 Nucl. Instr. Meth. A, 436, 91
%The X-ray CCD Camera of the MAXI Experiment on the JEM/SSC 
%     E.Miyata, H.Tsunemi, H.Ogata, D.Akutsu, K.Yoshita, Y.Hashimoto, K.Torii,M.Matsuoka, N.Kawai,
%     A.Toshida, T.Mihara, T.Kotani, H.Negoro, H.Kubo, H.Matsumoto, Y.Shirasaki, B.C.Rubin, I.Sakurai and
%     M.Yamauchi 
%     Nucl. Instr. Method A, 1999, 436, 91-95 

\re
Miyata E. et al. 2000 proc of SPIE (New Astronomy 2000), 4012, 186
% Performance and calibration of the x-ray CCD camera of the MAXI experiment on the ISS/JEM
% E.Miyata, C.Natsukari,D.Akutsu, M.Ohtani,H.Tsunemi, M.Matsuoka, N.Kawai,

\re
Tomida H. et al. 1997 PASJ., 49, 389

\re
Tomida H. et al. 2000a proc of SPIE (New Astronomy 2000), 4012, 178
%The MAXI mission on the International Space Station 
%     H.Tomida, M.Matsuoka, S.Ueno, K.Torii, M.Sugizaki, W.M.Yuan, S.Komatsu, Y.Shirasaki, N.Kawai,
%     S.Yoshida, T.Mihara, I.Sakurai, H.Negoro, H,Tsunemi, E.Miyata, M.Yamauchi, I.Tanaka, 
%     SPIE(New Astronomy2000 : X-Ray Optics, Instruments, and Missions III), 2000, 4012, 178

\re
Tomida H. et al. 2000b proc of SPIE., 4140, 304
%Solid state Slit Camera (SSC) of the MAXI Mission for the JEM (Japanese Experiment Module) on the International Space Station (ISS) 
%     H.Tomida, M.Matsuoka, K.Torii, S.Ueno, M.Sugizaki, W.M.Yuan, Y.Shirasaki, M.Sakano, S.Komatsu,
%     H.Tsunemi, E.Miyata, N.Kawai, A.Yoshida, T.Mihara, I.Tanaka 
%     SPIE, 2000,4140, 304

\re
Torii K. et al. 1999 proc of SPIE., 3765, 636
%X-ray detectors and calibration system for the MAXI mission 
%     K.Torii, M.Matsuoka,M.Sugizaki, H.Tomida, S.Ueno,W.M.Yuan, N.Kawai, T.Mihara, A.Yoshida,
%     H.Tsunemi, E.Miyata, H.Negoro, I.Sakurai, Y.Shirasaki, D. Akutsu, K.Hayashida, 
%     SPIE, 1999, 3765, p636

\re
Yamashita A. et al. 1997 IEEE Trans. Nucl. Sci., 44, 847
% \bibitem{Yamashita1997}
% A.~{Yamashita}, T.~{Dotani}, M.~{Bautz}, G.~{Crew}, H.~{Ezuka}, K.~{Gendreau},
%   T.~{Kotani}, K.~{Mitsuda}, C.~{Otani}, A.~{Rasmussen}, G.~{Ricker}, and
%   H.~{Tsunemi}.
% \newblock {\em IEEE Trans. Nucl. Sci.}, 44:847, 1997.

\re
Yuan W. et al. 2001 in this proceeding

\end{document}